\documentclass[aps,amsmath,reprint, superscriptaddress, amssymb, nofootinbib]{revtex4-2}
\usepackage{graphicx}
\usepackage[utf8]{inputenc}
\usepackage[T1]{fontenc}
\usepackage{lmodern}
\usepackage{dcolumn}
\usepackage{bm}
\usepackage{amsmath}
\usepackage[dvipsnames,x11names]{xcolor}

\definecolor{newgreen}{rgb}{0.01,.75, 0.24}

\newcommand{\ord}[1]{\mathcal{O}\left( #1 \right)}

\begin{document}

  \title{Initial-State Charge Density Predicts Final-State Net Charge Flow in Heavy-Ion Collisions}
\author{Jefferson Sousa}
\affiliation{
Universidade de Sao Paulo, Instituto de Fisica,
Rua do Matao 1371,
05508-090
Sao Paulo, SP, Brazil
}

\author{Matthew D. Sievert}
\email[Email: ]{msievert@nmsu.edu}
\affiliation{Department of Physics, New Mexico State University, Las Cruces, NM 88003, USA}
\author{Jacquelyn Noronha-Hostler}
\email[Email: ]{jnorhos@illinois.edu}
\affiliation{The Grainger College of Engineering, Illinois Center for Advanced Studies of the Universe, Department of Physics, University of Illinois at Urbana-Champaign, Urbana, IL 61801, USA}
\author{Patrick Carzon}
\email[Email: ]{pcarzon@franciscan.edu}
\affiliation{Department of Mathematics and Physical Sciences, Franciscan University of Steubenville, Steubenville, OH 43952 }
%
\author{Matthew Luzum} 
\email[Email: ]{mluzum@usp.br}
\affiliation{
Universidade de Sao Paulo, Instituto de Fisica,
Rua do Matao 1371,
05508-090
Sao Paulo, SP, Brazil
}
\begin{abstract}
We propose a new class of charge-conjugation-odd flow observables and use them to investigate the dynamics of conserved currents in simulations of relativistic heavy-ion collisions.
Inspired by the success of the initial energy and momentum distributions at predicting final-state anisotropic flow, we construct systematically-improvable initial-state estimators for final net-charge flow observables, which we validate with numerical simulations. This opens the possibility of a multitude of new charge-dependent probes of heavy-ion collisions of different systems and energies.

\end{abstract}

\maketitle

\section{Introduction}
Over the last decades, a robust paradigm has emerged about the mapping between the initial state produced in ultra-relativistic collisions of heavy ions at facilities like the Relativistic Heavy Ion Collider (RHIC) and the Large Hadron Collider (LHC) and the final-state distribution of particles which are directly detected.  In this paradigm, while the initial state is not directly observed, it can be characterized by the spatial distribution of the energy density (or an equivalent thermodynamic quantity, like the entropy density).  This initial geometry can be inferred indirectly, through knowledge of the system's hydrodynamic evolution during the formation and extinction of the quark-gluon plasma (QGP).

Quantifiers of the initial-state energy geometry, the traditional eccentricity vectors $\mathcal{E}_n$, have been shown to serve as accurate predictors of event-by-event final-state flow $V_n$.
The full complexity of the QGP and its hydrodynamic response to the initial-state geometry can be encoded remarkably well in a few coefficients describing the linear (and higher) response of the flow vectors $V_n$ to the initial-state geometry $\mathcal{E}_n$\cite{Teaney:2010vd,Gardim:2011xv,Noronha-Hostler:2015dbi,Sievert:2019zjr,Rao:2019vgy}. 

These eccentricities $\mathcal{E}_n$ arise naturally as the longest-wavelength modes of the initial state with rotational symmetry commensurate with $V_n$.  The traditional eccentricities therefore belong to a family of systematically improvable initial-state estimators 
$\mathcal{E}_{n,m}$ 
with the same Fourier harmonic $n$, but at different length scales $m \geq n$, with $\mathcal{E}_n = \mathcal{E}_{n,n}$.  
Recently, this paradigm was extended to encompass other components of the initial energy-momentum tensor: the initial momentum density $T^{0i}$ and the initial stress $T^{ij}$.  While initial momentum degrees of freedom have a small impact in large systems like PbPb, they can dominate over the initial energy eccentricities in small systems \cite{Schenke:2019pmk}.  From PbPb to pPb to pp, the linear correlation coefficient $Q_2$ of the elliptic flow to the initial-state energy eccentricity falls dramatically.  
However, the predictive power of the initial state in small systems is  saved by the inclusion of the initial momentum degrees of freedom $T^{0i}$ and $T^{ij}$, which restore linear response 
across all three systems \cite{Sousa:2024msh}.

The striking success of the initial-state estimator paradigm and simple linear (and higher) response illustrates the power of their organizing principles.  The estimators are organized around a hierarchy of transverse length scales, under the principle that observables that survive after hydrodynamic evolution are more sensitive to long-wavelength features of the initial state.  The other guiding principle is symmetry matching: the eccentricity $\mathcal{E}_n$ has the same Fourier harmonic as $V_n$ in order to match its transformation law under rotational symmetry.  A similar matching of rotational modes governs the mixing of angular harmonics permitted in nonlinear response to the initial state.  The generality of these symmetry principles are what underpins the predictive power of the initial state across a host of collision systems.

What is currently missing from the field is a further generalization of the initial-state estimator framework beyond the energy-momentum tensor entirely, to encompass the conserved charges in the system such as baryon number $B$, strangeness $S$, and electric charge $Q$.  Much of the recent interest in conserved charges in heavy-ion collisions has focused on the regime of low center-of-mass energy $\sqrt{s}$, where the nuclei collide on long enough time scales to allow for significant baryon stopping, leading to a finite net baryon number $n_B>0$. A number of recent relativistic viscous hydrodynamic calculations have begun to explore this regime \cite{Denicol:2018wdp,Inghirami:2022afu,Schafer:2021csj,Du:2024pbd}, comparing to collective flow data at low $\sqrt{s}$.  While it is understood that baryon diffusion plays a role \cite{Denicol:2018wdp,Du:2024pbd}, especially for observables that compare the difference between protons and anti-protons, the mapping between the initial state with stopped baryons and the final flow harmonics is unknown.  Further complicating the study of charge transport at low $\sqrt{s}$ are the far-from-equilibrium effects anticipated at the quantum chromodynamics (QCD) critical point \cite{Monnai:2016kud,Rajagopal:2019xwg,Dore:2020jye,Dore:2022qyz,Du:2021zqz,Pradeep:2024cca,Chattopadhyay:2024bcv}.  It is not yet clear  what role critical fluctuations and dynamical critical scaling of transport coefficients will play in dynamical simulations, since  full relativistic viscous hydrodynamic simulations with critical fluctuations are still a work in progress.  Constraining the behavior of charge transport in relativistic hydrodynamics \textit{absent} these critical phenomena will provide an important baseline to isolate the signatures of the critical point.

Furthermore, even at high $\sqrt{s}$ where the \textit{net} baryon stopping vanishes and the plasma is formed from the neutral gluon fields of the colliding nuclei, the perturbative splittings of these gluons into quark/anti-quark pairs can produce \textit{fluctuations} of all three BSQ charges \cite{Carzon:2019qja,Martinez:2019jbu} that in turn affect flow observables of identified particles \cite{Plumberg:2024leb,Gardim:2024nyz}.  Charge-rich plasmas can also be studied in high-energy collisions at far forward rapidity, for instance in the Color-Glass Condensate model \cite{Garcia-Montero:2023gex}.  If charm quarks are thermalized within such high-temperature quark-gluon plasmas, they could provide a fourth conserved charge $BSQC$ within the fluid \cite{Capellino:2022nvf,Capellino:2023cxe}.

In this Letter, we present a new set of initial-state estimators sensitive to the initial density of one or more conserved charges.  These estimators are manifestly \textit{antisymmetric} under discrete symmetries such as charge conjugation $\mathcal{C}$.  We further introduce the commensurate antisymmetric final-state flow observables, which we refer to as ``net flow'' vectors $V_n^\mathrm{net}$.  We use a multi-stage simulation model utilizing a fluctuating Glauber initial condition (superMC+MUSIC+UrQMD) \cite{Shen:2020jwv, superMC,Shen:2014vra} to test our approach  and show decisively that the event-by-event net-proton elliptic flow is well predicted by the initial baryon density distribution. Our formalism is dictated by general symmetry principles, and so can be used both at low $\sqrt{s}$ where the system has a nonzero net conserved charge and also at large $\sqrt{s}$ where only local fluctuations of conserved charges can occur.  Within this framework, the tremendously successful paradigm of hydrodynamics as a long-wavelength deterministic response to the initial-state geometry can be extended to a wealth of new charge-dependent correlations.

\section{Proposed net flow estimators}
Soft-sector observables in relativistic heavy-ion collisions are well described by relativistic hydrodynamics, and it is generally expected that charge-dependent evolution is subleading to the main flow of the overall energy and momentum of the system.   As such, it is advantageous to devise observables that isolate the flow of conserved charges.  The mechanism by which we can do so is a generalized charge conjugation: quantities which are \textit{antisymmetric} under such a discrete symmetry decouple from those which are \textit{symmetric} (independent of charge), like the overall energy-momentum flow.  
To that end, we define a family of charge-dependent flow coefficients which generalize the usual flow coefficients $V_n$ and separate them into components $V_n^+$, $V_n^\mathrm{net}$ which are even and odd under generalized charge conjugation, respectively.

Consider a set of hadrons $\{h\}$ with conserved charges $\{q\}$ (of any kind) and identified anisotropic flow coefficients $V_n^{h}$.  That is, we define the theoretical single-event flow coefficient of angular harmonic $n$ for particle type $h$ as $V_n^h = \langle e^{in\phi} \rangle_h$, where the average is over all particles of species $h$ in a given event.  Examples include flow coefficients for single particles like $V_n^{\pi^+}$, $V_n^{K^+}$ as well as linear combinations of particles $V_n^{p + K^+} = V_n^{p} + V_n^{K^+}$.  Then, by subtracting the event-by-event flow coefficients of the corresponding \textit{antiparticles}, we can construct \textit{net} charge flow coefficients $V_n^\mathrm{net}$ which are manifestly \textit{odd} under the conjugation of one or more conserved charges, as illustrated below. 
\renewcommand{\arraystretch}{1.5}
\begin{center}
\begin{tabular}{|c|c|}
    \hline
    $V_n^{\rm net}$ &
        Charge Conjugated
    \\ \hline \hline
    $V_n^{n} - V_n^{\bar{n}}$   &  
        B
    \\
    $V_n^{K^0} - V_n^{\bar{K}^0}$   &  
        S
    \\
    $V_n^{\pi^+} - V_n^{\pi^-}$   &  
        Q
    \\ \hline \hline
    $V_n^{p} - V_n^{\bar{p}}$   &  
        B, Q
    \\
    $V_n^{K^+} - V_n^{K^-}$   &  
        S, Q
    \\
    $V_n^{\Lambda^0} - V_n^{\bar{\Lambda}^0}$   &  
        B, S
    \\ \hline 
\end{tabular}
\renewcommand{\arraystretch}{1}
\end{center}
One may also in general form linear combinations of such charged flow coefficients, creating combinations like $V_n^{\pi^+} - V_n^{\pi^-} + V_n^{K^0} + V_n^{\bar{K}^0}$ which are \textit{even} under one charge (S) and odd under another (Q).  Most particle species for which particle identification is available will have multiple charges, but it is still theoretically useful to consider net flow coefficients from species like neutral kaons and neutrons which conjugate only a single B/S/Q charge.

From these constituents, various experimental flow observables can be constructed in the usual way as currently done for identified particle flow.   For example 
\begin{align}
    v_n^{\rm net}\{2\} \equiv \frac{\langle V_n^{\rm net} \, V_n^{*} \rangle}{\sqrt{\langle | V_n |^2\rangle}}.
\end{align}

To construct an estimator for the single-event net flow $V_n^{\rm net}$, we follow the same guidelines as previous successful estimators that have been used for unidentified particle flow.  That is, we are guided by the separation of length scales, such that the bulk evolution of the system and the final observables are more sensitive to large-scale structures in the initial state than to small-scale features.  

Estimators are constructed by decomposing a generating function into discrete cumulants, which represent particular rotational modes and which (via a Fourier transform) are ordered according to length scale, and arranged in a power series.   The leading order estimator is then proportional to the lowest cumulant.  In analogy with the well-known eccentricity estimators we can write
\begin{align}
V^{\rm net}_n \simeq V_n^{\rm est} = \kappa_n\mathcal{E}^{\rm net}_n(\gamma).
\end{align}
where the leading-order estimator is a ratio of cumulants:
\begin{align}
\mathcal{E}^{\rm net}_n(\gamma) = - \frac{W^{net}_{n,n}(\gamma)}{\Big|W^+_{0,2}\Big|^\frac{n}{2}}.
\end{align}
In this case, the numerator is an anti-symmetrized (charge conjugation odd) cumulant, while the denominator is a symmetrized cumulant that represents the size of the system, which we use as a normalization scale
\begin{align}   \label{e:Symmetrize}
    W_{n,m}^{\rm net} \equiv \frac{1}{2} \left( W_{n,m} (\gamma) - W_{n,m} (-\gamma) \right),\\
    W_{n,m}^+ \equiv \frac{1}{2} \left( W_{n,m} (\gamma) + W_{n,m} (-\gamma) \right).
\end{align}
For a single conjugated charge (herein considered baryon number $B$), the proposed generating function is
\begin{align}   
\rho(\vec x_\perp) \equiv \varepsilon(\vec x_\perp) - E_\perp  \gamma_B \, n_B(\vec x_\perp).\label{Eq::Generating function}
\end{align}
where $\varepsilon$ is the energy density, $n_B$ is the density of the conserved current (here $B$), and $\gamma_B$ is a constant. From this the cumulants are obtained from the 2D Fourier transform $\tilde\rho$,
\begin{align}   \label{e:genfn2}
    \ln\left[\tilde\rho (\vec k_\bot)\right] = \sum_{n, m \geq |n|} W_{n,m} k^m e^{-in\phi_k}.
\end{align}

The cumulants must be generated iteratively order-by-order.  The lowest cumulants for the general case of multiple conserved charges can be found in the End Matter, but for a single conserved charge with $n=2$, we need only the following, expanded for small $\gamma_B \ll 1$
\begin{align}
\label{e:Wnet}
W^{net}_{2,2} &= \frac{1}{8} \Big( \gamma_B \{ (z - \{z\}_\varepsilon)^2 \}_{n_B} - \gamma_B Q_B \{ (z - \{z\}_\varepsilon)^2 \}_\varepsilon \Big) \nonumber,\\
  W^+_{0,2} &=  -\frac{1}{4} \{ |z - \{z\}_\varepsilon|^2 \}_\varepsilon .
\end{align}
with
\begin{align}
    E_\perp &\equiv \int d^2 x_\bot \, \varepsilon (\vec{x}_\bot)    \\
    Q_B &\equiv \int d^2 x_\bot \, n_B (\vec{x}_\bot)    \\
    \{ f \}_\varepsilon &\equiv \frac{1}{E_\bot} 
    \int d^2 x_\bot \, f(\vec{x}_\bot) \: \varepsilon (\vec{x}_\bot)    \\
    \{ f \}_{n_B} &\equiv  
    \int d^2 x_\bot \, f(\vec{x}_\bot) \: n_B (\vec{x}_\bot)
\end{align}
This estimator can be systematically improved with non-linear contributions and higher cumulants (with $m>n$), but we focus here on the leading order result which should be the most useful.  A full derivation and explanation can be found in the End Matter.

We note that in addition to the usual system response coefficients $\kappa_n$, there is an additional response coefficient $\gamma_B$ associated with the conserved charge $B$.  However, while there is a different coefficient for each conserved quantity, we note that a single value of $\gamma$ should describe all harmonics, as was found for the contribution from initial momentum degrees of freedom \cite{Sousa:2024msh}.

\section{Numerical validation}
We validate our proposed framework by performing event-by-event hybrid-hydrodynamic simulations with nonzero, fluctuating baryon currents.   We then test on an event-by-event basis how well the proposed estimator $V_2^{\rm est}$ predicts the net proton elliptic flow vector $V_2^{\rm net} = V_2^p - V_2^{\bar p}$, as an experimentally-relevant proxy for net baryon flow.

Details of the simulations can be found in the End Matter, but we highlight here that model (superMC+MUSIC+UrQMD) \cite{Shen:2020jwv, superMC,Shen:2014vra} has a realistic  description of energy and baryon number deposition, including event-by-event fluctuations, which therefore serves for as a stringent test of our net charge estimators on an event-by-event basis.  However, the predictive power of the estimators presented here is not unique to this system.

The quality of the estimator can be quantified via its linear (Pearson) correlation with the quantity it is intended to estimate
\begin{align}
\label{eq:pearson}
    Q_n\left\{A_n, B_n\right\}&\equiv \frac{\langle A_n B_n^*\rangle}{\sqrt{\langle |A_n|^2\rangle \langle |B_n|^2\rangle }}.
\end{align}
Here $A_n, B_n$ are two complex numbers that represent a final-state observable and initial-state estimator.  So a value of $Q_n = 1$ indicates that the estimator gives a perfect prediction for the observable in every event, while a value of $Q_n = 0$ indicates no (linear) correlation and consequently an estimator with no predictive power.

\begin{figure}[]\centering
\includegraphics[width=\linewidth]{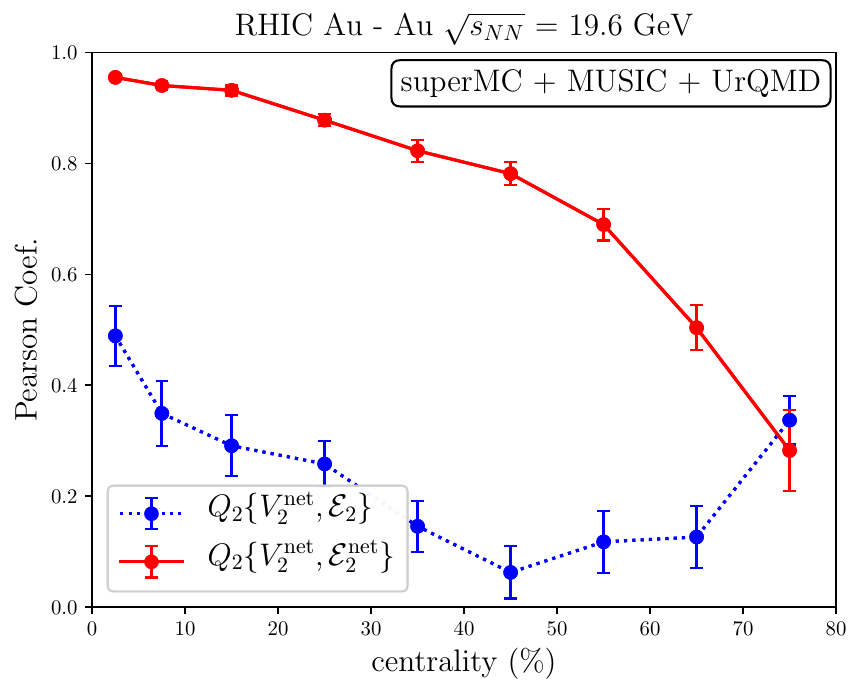}
\caption{
The dotted blue line represents the linear correlation coefficient \eqref{eq:pearson} between the net elliptical flow and the traditional eccentricity $Q_2\{V_2^{\rm net}, \mathcal{E}_2\}$, and the solid red line represents the linear correlation between the net elliptical flow and the net eccentricity $Q_2\{V_2^{\rm net}, \mathcal{E}_2^{\rm net}\}$.
\label{Fig::BaryonPearson}
}
\end{figure}

In Fig.~\ref{Fig::BaryonPearson} we show the resulting Pearson correlation coefficient $Q_2$ for two estimator/observable pairs.
Unsurprisingly, the usual eccentricity $\mathcal{E}_n$ is an exceptionally poor estimator for the \textit{net} elliptic flow $V_n^{\rm net}$, as can be seen by the blue curve of Fig.~\ref{Fig::BaryonPearson}, showing a very small $Q_2$.
The red curve, on the other hand, shows the Pearson correlation between the net elliptic flow and our proposed estimator.  Remarkably, the estimator shows excellent event-by-event performance, with a linear correlation greater than 0.9 for central and mid-central collisions.  The estimator performance decreases for more peripheral centralities, perhaps indicating expected contributions from non-linear corrections to the estimator.   However, we note that due to the small number of antiprotons, the statistical uncertainty becomes large and likely contributes to the small $Q_2$ in peripheral collisions.

\section{Conclusions}

We proposed a new class of conserved-charge-dependent flow observables that are antisymmetric with respect to one or more conserved quantities, designed to isolate charge-dependent effects.  To understand the dynamics that lead to these observables and make a direct connection to early-time densities, we developed a systematically-improvable framework for estimating the final net charge flow from initial-state densities.   We tested the leading-order estimators for net-baryon flow using event-by-event hybrid hydrodynamic simulations, showing remarkable predictive power and validating the proposed framework.  With these new tools for understanding conserved charge flow, combined with the large set of potential observables, new understanding can be gained from relativistic heavy-ion collisions.

\section*{Acknowledgments}
J.N.H. acknowledges support from the US-DOE Nuclear Science Grant No. DE-SC0023861.  M.S. acknowledges support from the US-DOE Nuclear Science Grant No. DE-SC0024560. J.N.H and M.S. acknowledge support within the framework of the Saturated Glue (SURGE) Topical Theory Collaboration. This work was also supported in part by the NSF under grant number OAC-2103680 within the MUSES Collaboration.  M.L.~was supported by FAPESP projects 2017/05685-2, 2018/24720-6, and 2023/13749-1, by project INCT-FNA Proc.~No.~464898/2014-5, and by CAPES - Finance Code 001.

\bibliography{inspire,NOTinspire}

\newpage
 
\section*{End Matter}

\subsection{Initial-state estimator framework}

We begin with the notion that large-scale features that are initially present in the system have a larger effect on final observables than small-scale features.  With this hierarchy of scales, we can construct estimators for final-state observables in the form of a series which can be truncated and/or improved by including more terms in the expansion.  As introduced in Eqs.~\eqref{Eq::Generating function} and \eqref{e:genfn2}, we define a generating function $\rho(\vec{x}_\bot)$ in transverse position space and its Fourier transform $\tilde\rho(\vec{k}_\bot)$ such that
\begin{align}   \label{e:End:genfn1}
    \exp\Big( W(\vec{k}_\bot) \Big) \equiv \int d^2 x_\bot \: e^{i \vec{k}_\bot \cdot \vec{x}_\bot} \: \rho(\vec{x}_\bot) \equiv \tilde\rho(\vec{k}_\bot)   \: ,
\end{align}
with cumulants $W_{n, m}$ defined by the Fourier-Taylor expansion
\begin{align}   \label{e:End:genfn2}
    W(\vec{k}_\bot) &\equiv \ln\left[ \tilde\rho\left( \vec{k}_\bot \right) \right] 
    \notag \\ &\equiv \sum_{m = 0}^\infty \sum_{n = - m}^m \: W_{n, m} \: k^m \: e^{- i n \phi_k}   \: ,
\end{align}
with $k, \phi_k$ the magnitude and angle of $\vec{k}_\bot$, respectively.  For a given angular harmonic $n$, the assumption of long-wavelength dominance in hydrodynamic response implies that the lowest values of the radial harmonic $m \rightarrow |n|$ dominate.  The cumulants 
\begin{align}   \label{e:End:genfn3}
    W_{n, m} = \int\frac{d\phi_k}{2\pi} \, e^{i n \phi_k} \: \frac{1}{m!} \left.\frac{\partial^m}{\partial k^m}\right|_{k=0} \: W(\vec{k}_\bot)
\end{align}
thus provide a complete and controlled expansion for the generating function $\rho(\vec{x}_\bot)$.  Note that $W_{-n, m} = (-1)^n \: W_{n, m}$ and that various conventions for $W_{n,m}$ exist in the literature.

In the conventional initial-state estimator paradigm, the generating function $\rho(\vec{x}_\bot)$ is taken to be the energy density $\varepsilon(\vec{x}_\bot) = T^{\tau \tau}(\vec{x}_\bot)$ (or an equivalent thermodynamic quantity, like the entropy density).  The key new idea introduced in this work is to extend the definition of the generating function to incorporate the effect of the charge density $n(\vec{x}_\bot) = J^\tau (\vec{x}_\bot)$.  In general, let $\mathcal{A}$ denote the set of conjugated charges over which we will antisymmetrize, and let $\mathcal{S}$ denote the set of remaining charges over which we will symmetrize.  We accordingly define the extended generating function $\rho(\vec{x}_\bot; \gamma)$ as in Eq.~\eqref{Eq::Generating function} as
\begin{align}
    \rho(\vec{x}_\bot; \gamma) \equiv \varepsilon (\vec{x}_\bot) - E_\bot \:
    \sum_i^{\mathcal{A},\mathcal{S}} \gamma_i \, n_i (\vec{x}_\bot)   \: ,
\end{align}
where the summation runs over all the various types of conserved charge (e.g., B, S, Q) in both $\mathcal{A}$ and $\mathcal{S}$, and $\gamma_i$ acts a coupling constant which determines the impact of charge type $i$ on the generating function.  Here $E_\bot \equiv \int d^2 x_\bot \: \varepsilon(\vec{x}_\bot)$ is the total energy.  Given the success of the energy-momentum-tensor-based estimators for describing $C$-even observables, evidently $\gamma \ll 1$; this is also consistent with the results of our simulations.

We may then proceed to straightforwardly evaluate the cumulants $W_{n,m} (\gamma)$ using \eqref{e:End:genfn3} order by order in $(n, m)$, expressing the results in terms of the total charge $Q_i = \int d^2 x_\bot \: n_i (\vec{x}_\bot)$.  We can do this compactly in terms of the average 
\begin{align}
    \langle f \rangle_\gamma \equiv 
    \frac{
    \int d^2 \mathbf{z} \: f(\mathbf{z}) \: \rho(\mathbf{z}; \gamma)
    }
    {
    \int d^2 \mathbf{z} \: \rho(\mathbf{z}; \gamma)
    }
    \: ,
\end{align}
where we have used boldfaced notation to denote the complex two-vector, $\mathbf{z} = x_\bot^1 + i x_\bot^2$: an equivalent representation of the usual Euclidean two-vector $\vec{x}_\bot = x_\bot^1 \hat{i} + x_\bot^2 \hat{j}$, as in \cite{Gronqvist:2016hym}.  We find the first few cumulants to be
\begin{subequations}
\begin{align}
    W_{0,0} (\gamma) &=
    \ln\left[ E_\bot \left(1 - \sum_i \gamma_i Q_i \right) \right]\: ,  \\
    W_{1,1}(\gamma) &= 
    \frac{i}{2} \{ \mathbf{z} \}_\gamma \: ,  \\
    W_{0,2}(\gamma) &= 
    \frac{-1}{4} \left\{
        \left| \mathbf{z} - \{ \mathbf{z} \}_\gamma \right|^2 
    \right\}_\gamma    \: ,    \\
    W_{2,2}(\gamma) &=
    \frac{-1}{4} \left\{
        \left( \mathbf{z} - \{ \mathbf{z} \}_\gamma \right)^2 
    \right\}_\gamma    \: ,    \\
    W_{1,3}(\gamma) &=
    \frac{-i}{16} \left\{
        \left( \mathbf{z} - \{ \mathbf{z} \}_\gamma \right) 
        \left| \mathbf{z} - \{ \mathbf{z} \}_\gamma \right|^2 
    \right\}_\gamma    \: ,    \\
    W_{3,3}(\gamma) &=
    \frac{-i}{48} \left\{
        \left( \mathbf{z} - \{ \mathbf{z} \}_\gamma \right)^3 
    \right\}_\gamma    \: .
\end{align}
\end{subequations}
Note that for $n > 1$ the cumulants are translationally invariant (and thus good candidates for final-state flow estimators).  

We then form antisymmetric (symmetric) cumulants $W_{n,m}^{net}$ ($W_{n,m}^+)$ under the conjugation $\gamma_j \rightarrow - \gamma_j$ of at least one charge $n_j$.  We will compactly denote averaging of a function $f(\vec{x}_\bot)$ in the conjugated density as $\{ f \}_{\bar{\gamma}}$.

The directed $(1,1)$ net cumulant is just proportional to the dipole moment:
\begin{align}
    W_{1,1}^{net} &\equiv \frac{i}{4} \Big( \{ \mathbf{z} \}_\gamma - 
    \{ \mathbf{z} \}_{\bar\gamma} \Big)
    \notag \\ &=
    \sum_{j}^{\mathcal{A}} \frac{i}{2} \, \gamma_j \, \{ \mathbf{z} \}_{n_j} + \ord{\gamma^3}    \: .
\end{align}
The elliptic $(2,2)$ net cumulant is a combination of the ellipticity of the energy density, weighted by the charges, and the ellipticity of the charge distribution:
\begin{align}
    W_{2,2}^{net} &\equiv \frac{-1}{8} \Big( \left\{
        \left( \mathbf{z} - \{ \mathbf{z} \}_\gamma \right)^2 \right\}_\gamma 
        - \left\{\left( \mathbf{z} - \{ \mathbf{z} \}_{\bar\gamma} \right)^2 
    \right\}_{\bar\gamma } \Big)
    \notag \\ &=
    \sum_{j}^{\mathcal{A}} \frac{1}{8} \Big( \gamma_j \{ (\mathbf{z} - \{z\}_\varepsilon)^2 \}_{n_j} - \gamma_j Q_j \{ (\mathbf{z} - \{z\}_\varepsilon)^2 \}_\varepsilon \Big)
    \notag \\ &\hspace{1cm} + \ord{\gamma^3} \: .
\end{align}
Finally, the triangular $(3,3)$ net moment includes not only a triangular analog of $W_{2,2}^{net}$, but also a cross term which couples the $(1,1)$ cumulants with the $(2,2)$ cumulants:
\begin{align}
    W_{3,3}^{net} &\equiv \frac{-i}{96} \Big( \left\{
        \left( \mathbf{z} - \{ \mathbf{z} \}_\gamma \right)^3 \right\}_\gamma 
        - \left\{\left( \mathbf{z} - \{ \mathbf{z} \}_{\bar\gamma} \right)^3 
    \right\}_{\bar\gamma } \Big)
    \notag \\ &=
    \sum_{j}^{\mathcal{A}} \frac{i}{48} \Big[ 
        \gamma_j \{ (\mathbf{z} - \{z\}_\varepsilon)^3 \}_{n_j} 
        - \gamma_j Q_j \{ (\mathbf{z} - \{z\}_\varepsilon)^3 \}_\varepsilon 
    \notag \\ &
        - 3 \Big( \gamma_j \{ \mathbf{z} \}_{n_j} - \gamma_j Q_j \, \{\mathbf{z}\}_\varepsilon \Big) \{ (\mathbf{z} - \{z\}_\varepsilon)^2 \}_\varepsilon  \Big]
    \notag \\ \hspace{1cm} &
        + \ord{\gamma^3} \: .
\end{align}
In this way one may systematically compute all the net cumulants, as well as the symmetric versions if desired.  Note, however, that the symmetric part is dominated by the energy density $\varepsilon$, with corrections being suppressed by the empirically small parameter $\gamma \ll 1$.  As such, corrections to the symmetric part enter at higher orders in $\gamma$ and are not shown explicitly.  For example, in this limit the symmetrized charge-dependent cumulant $W_{2,0}^+$ reduces back to the ordinary elliptic cumulant of the energy density:
\begin{align}
    W_{0,2}^{+} &\equiv \frac{-1}{8} \Big( \left\{
        \left| \mathbf{z} - \{ \mathbf{z} \}_\gamma \right|^2 \right\}_\gamma 
        + \left\{\left| \mathbf{z} - \{ \mathbf{z} \}_{\bar\gamma} \right|^2 
    \right\}_{\bar\gamma } \Big)
    \notag \\ &=
    \frac{-1}{4} \{ |\mathbf{z} - \{\mathbf{z}\}_\varepsilon|^2 \}_{\varepsilon} + \ord{\gamma^2} \: .
\end{align}
Due to the smallness of $\gamma$, which in this work is $\gamma \sim \ord{10\%}$, the corrections which are suppressed by a relative factor of $\gamma^2 \sim \ord{1\%}$ are negligible, and the linearized dependence on $\gamma$ is adequate for our purposes.

\subsection{Hybrid simulations}

The numerical simulations were performed with the multi-stage iEBE-MUSIC simulation chain \cite{superMC}.   The initial conditions are given by a Monte Carlo Glauber model which is followed by relativistic viscous hydrodynamic evolution, which is then followed by a UrQMD hadron cascade afterburner.  2250 hydrodynamic events were simulated, with 10000 UrQMD afterburner events for each hydrodynamic event.  
The Glauber model invokes a string picture to generate a 3D, fluctuating, distribution of initial energy, ensuring global energy conservation.   Details of the model can be found in Ref.~\cite{Shen:2020jwv}.

The hydrodynamic evolution modeled by the DNMR equations of motion \cite{Denicol:2012cn},  with temperature-dependent shear and bulk viscosity where their parameterization comes from \cite{Bernhard:2018hnz} (we assume no dependence of viscosity on $\mu_B$), and a lattice-QCD-based equation of state (EoS) at finite net baryon density, NEoS-BQS \cite{Monnai:2019hkn}.   The NEoS-BQS EoS imposes strangeness neutrality and
constrains the local net electric charge density to be 0.4
times of the local net baryon density, which is reasonable for gold-gold collisions.

\begin{figure}[h]\centering
\includegraphics[scale=0.6]{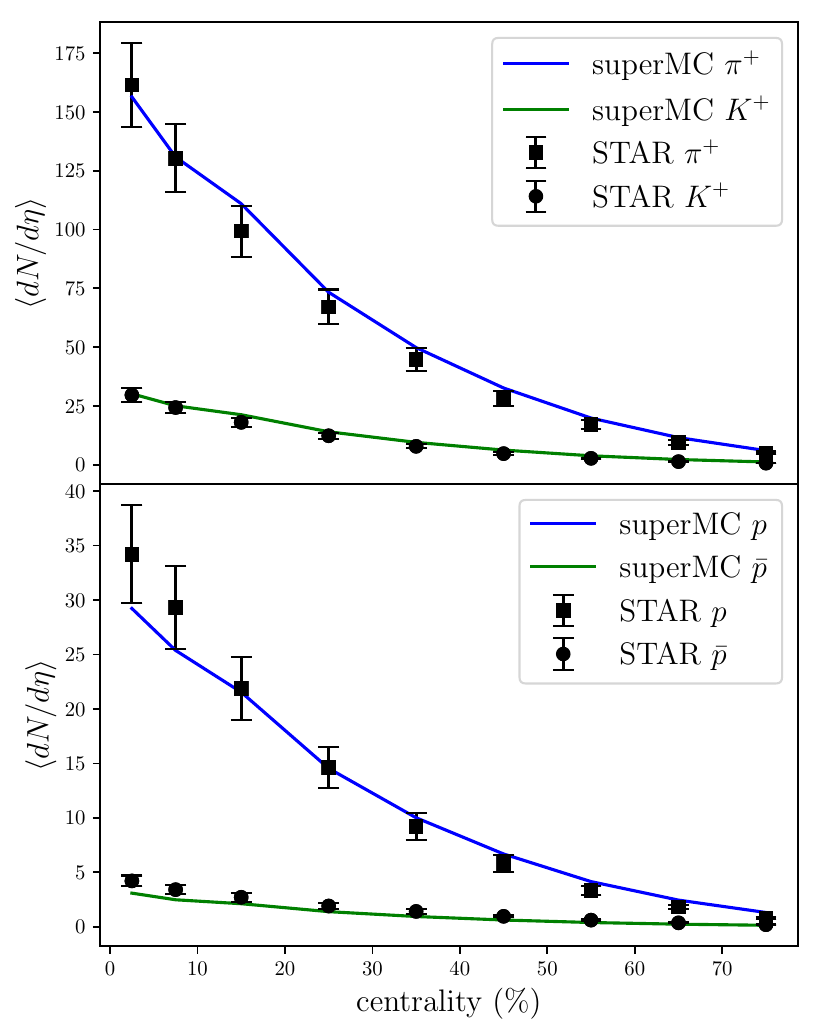}
\caption{Identified particle yields in 19.6 GeV Au-Au collisions.  Shown are experimental measurements from the STAR Collaboration \cite{STAR:2017sal} compared to simulations from this work.}
\label{Fig::BaryonMultiplicity}
\end{figure}

The hydrodynamic parameters were set according to a previous Bayesian analysis \cite{Bernhard:2018hnz}, while
initial-state parameters were tuned to obtain a good description of 19.6 TeV Au+Au identified particle yields as measured by the STAR collaboration \cite{STAR:2017sal}.  The parameter values are listed in Table \ref{Tab::superMC}, and the resulting particle yields are shown in Fig.~\ref{Fig::BaryonMultiplicity}.

\begin{table}[h]\centering
\caption{Parameters used for constructing the superMC (Glauber) initial conditions for Au-Au Collisions at 19.6 GeV \cite{Shen:2020jwv}.
}
\begin{tabular}{|c|c|c|c|c|c|c|c|} \hline
$\sqrt{s_{NN}}$ (GeV) & $\tau_0$ (fm) & $s_0$ & $\eta^s_0$ & $\sigma_{\eta,s}$ & $\eta^{n_B}_B$ & $\sigma_{\eta,+}$ & $\sigma_{\eta,-}$\\ \hline\hline
$19.6$ & $1.5$ & $6.3$ & $2.7$ & $0.3$ & $1.5$ & $0.2$ & $1.0$\\ \hline
\end{tabular}
\label{Tab::superMC}
\end{table}

After hydrodynamic evolution, the fluid degrees of freedom are converted to hadronic distribution functions via the usual Cooper-Frye procedure.   Then the hadrons and resonances are evolved according to the Boltzmann equation via the UrQMD hadron cascade \cite{Bleicher:1999xi}.  UrQMD works via a finite sampling of discrete particles, but in order to obtain an accurate value for the single-event net proton elliptic flow $V_2^{net}$ we sample many UrQMD events for each hydrodynamic simulation to reconstruct the relevant single-body distribution function.

\begin{figure}[!h]
\centering
\includegraphics[width=\linewidth]{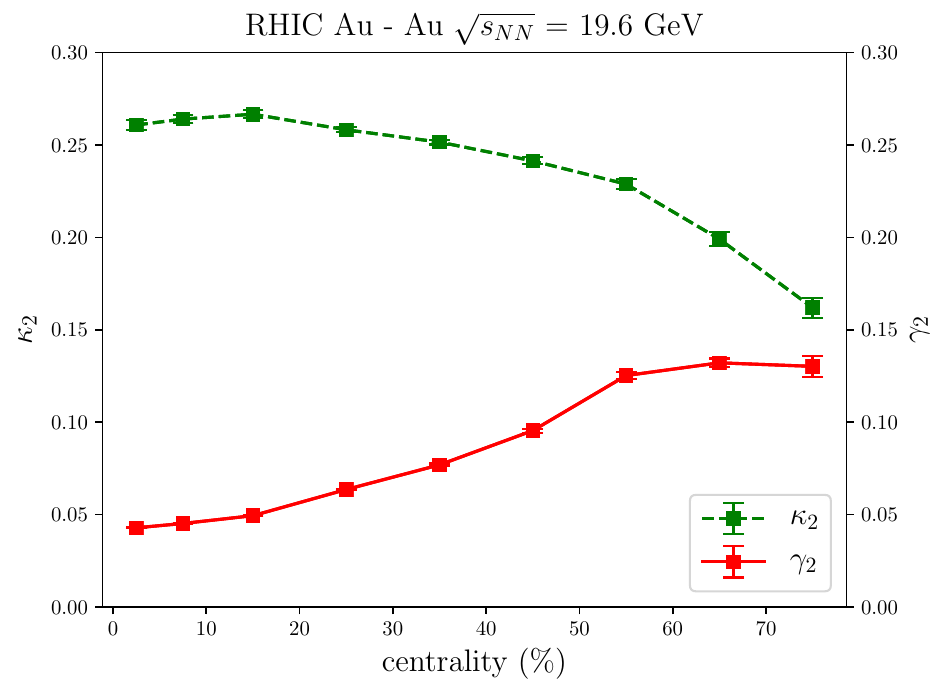}
\caption{The dashed green line represents the response coefficient $\kappa_2$, while the solid red line represents the response coefficient $\gamma_2$.
}
\label{Fig::BaryonResoponse}
\end{figure}

We also compute the values of the response coefficients $\kappa_2$ and $\gamma_B$.   We first note that the estimator (written to first order in $\gamma_B$, Eq.~\eqref{e:Wnet}) is proportional to $\kappa_2\gamma_B$, and so one cannot determine both coefficients from net proton flow alone.   Instead we determine $\kappa_2$  (the response to the initial energy distribution) in the usual way from charged hadron elliptic flow.   Then $\kappa_2\gamma_B$ (and therefore $\gamma_B$) is determined by maximizing the effectiveness of the net proton flow estimator.  Specifically, a maximal value of the Pearson coefficient between eccentricity and charged hadron elliptic flow is obtained by choosing 
\begin{equation}
   \kappa_2 = \frac{\mathrm{Re}\langle V_2 \, \mathcal{E}_2\rangle}{\langle|\mathcal{E}_2|^2\rangle}.
\end{equation}\\
\noindent{} Then, the optimal value of $\gamma_B$ becomes
\begin{equation}
    \gamma_B = \frac{\mathrm{Re}\langle V^{\rm net}_2\mathcal{E}^{\rm net}_2\rangle}{\kappa_2\langle|\mathcal{E}^{\rm net}_2|^2\rangle}.
\end{equation}
The extracted response coefficients are shown in Fig.~\ref{Fig::BaryonResoponse}.  We can see that the relative importance of the initial baryon density to net-proton elliptic flow increases for more peripheral centralities, possibly because of decreasing contribution from the energy density.

\end{document}